\documentclass[aps,pra,twocolumn,superscriptaddress]{revtex4-2}
\usepackage{amsmath,amsfonts,amssymb}
\usepackage{graphicx,float,calc}
\usepackage{color,bm}
\usepackage{ulem}
\usepackage{braket}
\usepackage[colorlinks,urlcolor=blue,citecolor=blue,linkcolor=blue]{hyperref}

\begin{document}

\title{Reentrant topological phases and spin density wave induced by 1D moir\'e potentials}

\author{Guo-Qing Zhang}
\affiliation{Research Center for Quantum Physics, Huzhou University, Huzhou 313000, P. R. China}

\author{Ling-Zhi Tang}
\email{tanglingzhi@quantumsc.cn}
\affiliation{Quantum Science Center of Guangdong-Hong Kong-Macao Greater Bay Area (Guangdong), Shenzhen 518045, China}

\author{L. F. Quezada}
\affiliation{Research Center for Quantum Physics, Huzhou University, Huzhou 313000, P. R. China}
\affiliation{Laboratorio de Ciencias de la Informaci\'{o}n Cu\'{a}ntica, Centro de Investigaci\'{o}n en Computaci\'{o}n, Instituto Polit\'{e}cnico Nacional, UPALM, 07700, Ciudad de M\'exico, M\'exico}

\author{Shi-Hai Dong}
\email{dongsh2@yahoo.com}
\affiliation{Research Center for Quantum Physics, Huzhou University, Huzhou 313000, P. R. China}
\affiliation{Laboratorio de Ciencias de la Informaci\'{o}n Cu\'{a}ntica, Centro de Investigaci\'{o}n en Computaci\'{o}n, Instituto Polit\'{e}cnico Nacional, UPALM, 07700, Ciudad de M\'exico, M\'exico}

\author{Dan-Wei Zhang}
\email{danweizhang@m.scnu.edu.cn}
\affiliation{Key Laboratory of Atomic and Subatomic Structure and Quantum Control (Ministry of Education), Guangdong Basic Research Center of Excellence for Structure and Fundamental Interactions of Matter, South China Normal University, Guangzhou 510006, China}
\affiliation{Guangdong Provincial Key Laboratory of Quantum Engineering and Quantum Materials, School of Physics, South China Normal University, Guangzhou 510006, China}

\date{\today}

\begin{abstract}Recent studies of 2D moir\'e  materials have opened opportunities for advancing condensed matter physics. However, the effect of 1D moir\'e  potentials on topological and correlated phases remains largely unexplored. Here we reveal a sequence of trivial-to-topological transitions and periodic-moir\'e-spin density waves induced by the 1D commensurate moir\'e potentials for spin-1/2 fermionic atoms. Such reentrant topology from a trivial phase is absent without the moir\'e potential and can be understood as the renormalization of topological parameters by the moir\'e strength. We then unveil the critical exponent and localization properties of the single-particle eigenstates. The periodic spin density wave of many-body ground states is contributed by the moir\'e potential, and is enhanced by on-site interactions but suppressed by nearest-neighbor interactions. Our results enrich the topological physics with multiple transitions and spin-density orders in 1D moir\'e systems, and the realization of the proposed model is promising in near-future ultracold atom setups.
\end{abstract}

\maketitle

\section{Introduction}

Topological phases have emerged as an important research field in condensed-matter physics~\cite{RevModPhys.82.3045,RevModPhys.83.1057,Culcer_2020,RevModPhys.88.021004,RevModPhys.88.035005,RevModPhys.90.015001} and engineered artificial systems~\cite{lu2014topological,RevModPhys.91.015006,Zhang2018,coen2024nonlinear,PhysRevLett.113.050402,roushan2014observation,PhysRevLett.120.130503,PhysRevLett.122.210401,PhysRevLett.130.036202,Li2024,Zhang2016}. These phases are typically protected by certain symmetries and characterized by topological invariants and non-trivial boundary modes, making them robust against local perturbations. Interactions can further enrich topological phases, giving rise to phenomena such as fractional quantum Hall effects~\cite{PhysRevLett.48.1559} and topological Mott insulators~\cite{PhysRevLett.100.156401,PhysRevA.86.053618,kuno2017various,PhysRevLett.110.260405,PhysRevB.88.045110,PhysRevA.101.013627,PhysRevB.101.235150,PhysRevB.99.075204,PhysRevB.104.L161118}. These topological properties have potential applications in fault-tolerant quantum computation~\cite{RevModPhys.80.1083,PoulsenNautrup2017,he2019topological} and spintronics devices~\cite{breunig2022opportunities,Ou2022,sahu2023room}. Quantum simulations of topological phases have achieved great progress, such as the realizations of the 1D topological insulator model~\cite{PhysRevLett.42.1698,PhysRevLett.110.076401,Song2018,Zhou2017}, 2D Harper-Hofstadter ~\cite{harper1955single} and Haldane models~\cite{PhysRevLett.61.2015}, and the observations of chiral edge states~\cite{doi:10.1126/science.aaa8515,doi:10.1126/science.aaa8736} with ultracold atoms in optical lattices.

In recent years, exotic properties have been unveiled in 2D moir\'e systems, including flat bands~\cite{cao2018correlated,PhysRevB.99.075127,balents2020superconductivity,haddadi2020moire,Hu2023,https://doi.org/10.1002/adma.202300572,Crepel2024}, moir\'e excitons~\cite{zhang2020twist,doi:10.1126/sciadv.abc5638}, interlayer ferromagnetism~\cite{doi:10.1126/science.aav1937}, and correlated topological states~\cite{dean2013hofstadter,PhysRevLett.122.016401,PhysRevB.99.205150,PhysRevResearch.3.L032070,doi:10.1126/science.aay5533,PhysRevB.108.155406}.  Due to the misalignment of two periodic lattices, moir\'e superlattices exhibit a new superlattice periodicity, which can be finite (commensurate) and infinite (incommensurate). Thus, the moir\'e systems provide a versatile platform for engineering band structures and correlated quantum states. In particular, the band width in moir\'e systems can be made extremely narrow, which enable to stabilize strongly-correlated superconductivity and topological insulators~\cite{cao2018unconventional,PhysRevLett.131.016001,nuckolls2020strongly,PhysRevB.107.165114,PhysRevLett.132.036501}. By reducing 2D twisted graphene to 1D incommensurate carbon nanotubes, the emerged moir\'e potentials significantly alter the band structure and characteristics of the system~\cite{PhysRevB.91.035405,PhysRevLett.124.106101}. Alternatively, the moir\'e analogs can be achieved by finite incommensurate 1D lattice potentials~\cite{PhysRevLett.126.036803,Goncalves2024} or finite coupled resonators with different modulation lengths~\cite{PhysRevLett.130.143801}. It has been revealed that the incommensurability can enable quasi-fractal charge-density waves (CDWs) in 1D narrow-band moir\'e systems~\cite{Goncalves2024}. The critical states induced by the incommensurability are theoretically predicted and experimentally observed~\cite{PhysRevLett.125.073204,PhysRevLett.131.176401,huang2025exactquantumcriticalstates_j}. The topological phase can be driven by incommensurate quasiperiodic disorders from a trivial phase \cite{PhysRevLett.125.073204,PhysRevLett.131.176401,huang2025exactquantumcriticalstates_j,PhysRevA.105.063327,Li2024}, similar to the topological Anderson insulators induced by random disorders \cite{PhysRevLett.102.136806,PhysRevLett.103.196805}. However, the topological properties and spin-density orders of many-body ground states in commensurable moir\'e systems remain largely unexplored. In particular, it is unclear whether the 1D commensurate moir\'e potentials can induce topological phases from a trivial phase with multiple transitions.

In this work, we address this question by exploring a 1D spin-1/2 fermionic optical lattice with a commensurate moir\'e potential and interatomic interactions. In the non-interacting limit, the system Hamiltonian is analytically solved in momentum space with reduced moir\'e Brillouin zones (MBZs) under periodic boundary conditions (PBCs) and numerically analyzed via the exact diagonalization (ED) method under open boundary conditions (OBCs). The topological characteristics of this 1D system are the winding number defined under the PBC~\cite{PhysRevLett.102.196804} and its generalization to the real space under the OBC~\cite{PhysRevLett.113.046802}. We find that the trivial band insulator can be driven to a nontrivial insulator by the moir\'e superlattice potential multiple times. This reentrant topological phase has a sequence of trivial-topological-trivial-topological-trivial transitions, which is revealed from the winding numbers, energy gaps and the edge states. For the topological transitions, we derive the scale invariance of the winding number with a critical exponent. The robustness of reentrant topology against disorders is also numerically demonstrated. By adopting the band flatness and the fractal dimension (FD), we study the localization of single-particle eigenstates induced by the moir\'e potential. In the interacting case, the moir\'e enlarged periodicity makes the system's Hilbert space computationally challenging in the ED. Thus, we use the density matrix renormalization group (DMRG) method to investigate the density wave orders and topology of the many-body ground state. We unveil the occurrence of periodic-moir\'e-spin density wave (PM-SDW) whenever the moir\'e potential is activated, in both non-interacting and interacting cases. The PM-SDW order is enhanced by the on-site interaction but suppressed by the nearest-neighbor interaction. In the presence of strong nearest-neighbor interaction, the CDW order with the moir\'e periodicity can emerge. Moreover, the multiple topological transitions are preserved in the interacting case. The reentrant topological phase is characterized by the many-body Berry phase and nontrivial edge excitations, and even hosts the SDW order, which corresponds to the topological PM-SDW. Finally, we show that the reentrant topological transition is due to the renormalization of the Zeeman strength by the moir\'e modulation, which is thus a generic phenomenon in 1D commensurable moir\'e systems.

\section{Results}
\subsection{\label{sec2}Model Hamiltonian}

We start by considering a 1D Raman optical lattice loaded with ultracold fermionic atoms \cite{PhysRevLett.110.076401,Song2018,Zhou2017}.
Two internal states of atoms are used to encode a spin-1/2 degree of freedom, and proper Raman laser beams can be used to engineer atomic hopping terms with effective spin-orbit couplings and Zeeman potentials. The tight-binding Hamiltonian of the system reads
\begin{equation}\begin{split}\label{Ham}
\hat{H}=&\hat{H}_0+\hat{H}_{\text{int}}\\
\hat{H}_0=&-t\sum_{j}(\hat{c}_{j,\uparrow}^{\dagger} \hat{c}_{j+1,\uparrow}-\hat{c}_{j,\downarrow}^{\dagger} \hat{c}_{j+1,\downarrow}+\mathrm{H.c.})\\
&+t_s\sum_{j}(\hat{c}_{j,\uparrow}^{\dagger} \hat{c}_{j+1,\downarrow}-\hat{c}_{j,\downarrow}^{\dagger} \hat{c}_{j+1 ,\uparrow}+\mathrm{H.c.})\\
&+\sum_{j} (m_z+m_{o,j})(\hat{n}_{j,\uparrow}-\hat{n}_{j,\downarrow}),\\
\hat{H}_{\text{int}}=&U\sum_{j}\hat{n}_{j,\uparrow} \hat{n}_{j, \downarrow}+V\sum_{j,\sigma}\hat{n}_{j,\sigma} \hat{n}_{j+1,\sigma}.
\end{split}\end{equation}
Here $\hat{c}_{j,\sigma}^{\dagger}$ creates a fermionic atom with spin-$\sigma$ ($\sigma=\uparrow,\downarrow$) on lattice site $j$, and $\hat{n}_{j,\sigma}=\hat{c}_{j,\sigma}^{\dagger}\hat{c}_{j ,\sigma}$ is the particle number operator. The parameters $t$ and $t_s$ denote the spin-dependent and spin-flip hopping strengths, respectively. The former term generates an effective phase and momentum kick to the atoms while preserving their spin, whereas the latter term can flip their spin. The Zeeman potential contains two parts, the uniform part with strength $m_z$ and the spatially modulated part $m_{o,j}$. We consider the modulation with a moir\'e potential
\begin{equation}
m_{o,j}=m_o[\cos(2\pi j/{a}_{1})+\cos(2\pi j/{a}_{2})],
\end{equation}
which is formed by the superposition of two periodic potentials of period ${a}_{\alpha}$ ($\alpha=1,2$). Here, we focus on the commensurate moir\'e potential by choosing ${a}_{1}=3$ and ${a}_{2}=7$, such that the lattice system has a period of ${a}_{12}={a}_{1}{a}_{2}=21$ with each supercell consisting of $21$ sites. Note that other moir\'e potentials with different values of $\{a_1,a_2\}$ can be taken to present similar properties. In addition to the on-site atomic repulsion of strength $U$, we include the nearest-neighbour repulsion of strength $V$ in the interaction Hamiltonian $\hat{H}_{\text{int}}$ in Eq. (\ref{Ham}). We assume the lattice length to be $L={a}_{12}A=21A$ with $A$ supercells and focus on the system at (near) half filling with the particle number $N_f=L$. We set $t=1$ as the energy unit hereafter.

In the absence of interactions ($U=V=0$) and moir\'e potential ($m_o=0$), the single-particle Hamiltonian $\hat{H}_0$ reduces to the chiral AIII-class model with a topological insulator when $|m_z|<2t$~\cite{PhysRevLett.110.076401}, which has been proposed \cite{PhysRevLett.110.076401} and experimentally realized with ultracold fermions in the 1D Raman optical lattice \cite{Song2018}. In the experiment, the parameters $t_s$ and $m_z$ are independently tunable via the Rabi frequencies and the two-photon detuning of the Raman coupling \cite{Song2018}, respectively. The challenging aspect for the realization of the model Hamiltonian is properly engineering the moir\'e Zeeman potential $m_{o,j}$ by adding Raman beams (and the interatomic interactions \cite{PhysRevLett.110.076401,Zhou2017}). We also note that the results presented below can be realized in the system of sizes $L=42,84$, which are within the reach of current optical lattices.  Here a multi-site unit cell is essential to capture the moir\'e nature, resulting in a multi-band system. This large periodicity poses significant challenges for ultracold atoms in optical lattices as typically only few bands are controlled in current experiments. Thus, the current technology would not permit yet to directly observe of our findings. However, recent advancements, such as the manipulating orbital degrees of freedom~\cite{kiefer2023ultracold} and high-fidelity imaging (99.4\% fidelity in 2.4 $\mu$s)~\cite{su2025fast}, indicate the rapid evolution of experimental techniques. We anticipate that such observations will be feasible in the near future, positioning our theoretical analysis as a guide for upcoming experimental efforts.

The chiral symmetry for the single-particle Hamiltonian is given by $\hat{C}\hat{H}_0\hat{C}^{-1}=-\hat{H}_0$, where $\hat{C}=\mathbf{I}_L\otimes \hat{\sigma}_x$ is the chiral symmetry operator with $\mathbf{I}_L$ a $L$-rank identity matrix and $\hat{\sigma}_{x,y,z}$ Pauli matrices.
In the presence of the moir\'e potential, the chirality of $\hat{H}_0$ presences, the corresponding  momentum Hamiltonian can be derived. By taking the Fourier transformation under the PBC, we can transform the moir\'e Zeeman terms to the momentum space as
\begin{equation}\begin{split}
\sum_j\cos(\frac{2\pi j}{{a}_{\alpha}})\hat{n}_{j,\sigma}=\frac{1}{2}\sum_k\left(\hat{c}_{k,\sigma}^\dagger \hat{c}_{k+\frac{2\pi}{{a}_{\alpha}},\sigma} + \mathrm{H.c.} \right).
\end{split}\end{equation}
By constructing the Bloch basis $\ket{BK}=[\hat{c}_{k+0\pi/{a}_{12}\uparrow}\  \hat{c}_{k+2\pi/{a}_{12}\uparrow} \ \cdots\  \hat{c}_{k+0\pi/{a}_{12}\downarrow}\  \hat{c}_{k+2\pi/{a}_{12}\downarrow}\  \cdots]^\mathrm{T}$ in the reduced MBZ $k\in [0,2\pi/a_{12})$, we obtain the Bloch Hamiltonian of $\hat{H}_0$ as
\begin{equation}\begin{split}\label{Hamk}
\hat{H}_\mathrm{B}(k)=&\left[\hat{\sigma}_z\otimes(m_z-2t\mathbf{C})-\hat{\sigma}_y\otimes(2t_s\mathbf{S})\right]\\
&+ \hat{\sigma}_z\otimes\frac{m_o}{2}\left(\mathbf{E} +\mathbf{F} \right), \\
\end{split}\end{equation}
where $\mathbf{C}=\mathrm{diag}[\cdots,\cos(k+k_i),\cdots]$ and  $\mathbf{S}=\mathrm{diag}[\cdots,\sin(k+k_i),\cdots]$ are ${a}_{12}$-rank diagonal matrices with $k_i=2\pi(i-1)/a_{12}$ and $i=1,2,\cdots,a_{12}$. The matrix $\mathbf{E}$ ($\mathbf{F}$) is an ${a}_{12}\times {a}_{12}$ square matrix with its elements satisfying $\mathbf{E}(i,j)=\delta_{j,(i+a_2)\%a_{12}}+\delta_{j,(i+a_{12}-a_2)\%a_{12}}$ [$\mathbf{F}(i,j)=\delta_{j,(i+a_1)\%a_{12}}+\delta_{j,(i+a_{12}-a_1)\%a_{12}}$], where $\%a_{12}$ means $\mathrm{mod}\ a_{12}$. Under the moir\'e potential, the single-particle energy spectrum generally separates to $2{a}_{12}$ subbands.

\subsection{Reentrant topological phases}
\begin{figure}[t!]
\centering
\includegraphics[width=0.48\textwidth]{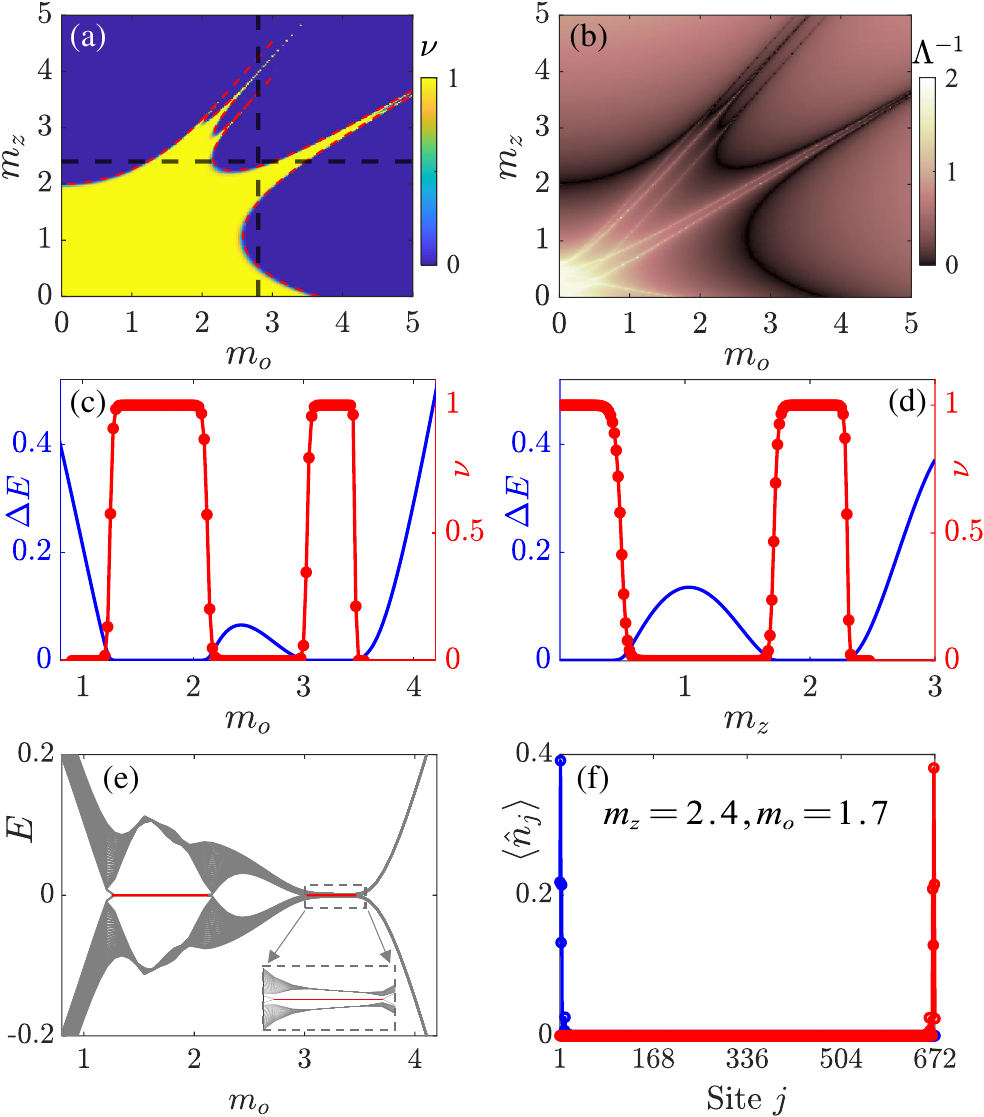}
\caption{Topological phase diagram and related properties in the single particle region. (a) Real-space winding number $\nu$ and (b) inverse of the zero-mode localization length $\Lambda^{-1}$ as functions of $m_o$ and $m_z$. Red dashed curve in (a) denotes the topological phase boundary revealed by the momentum-space winding number $\nu_k$. {\color{black}Horizontal and vertical black dashed lines in (a) correspond to the cuttings for $m_z=2.4$ in (c) and $m_o=2.8$ in (d), respectively.}
Real-space winding number $\nu$ and energy gap $\Delta E$ as functions of (c) $m_o$ for $m_z=2.4$ and (d) $m_z$ for $m_o=2.8$ under the OBC. (e) Energy spectrum with respect to $m_o$ under the OBC. The zero-energy modes in the topological regions are highlighted in red. The zoom in shows a detailed view of the energy spectrum in the second topological region.
(f) Density distributions of the two zero-energy edge modes for $m_z=2.4$ and $m_o=1.7$. Other parameters are $t_s=0.95$, and $A=32$.}
\label{fig1}
\end{figure}

The topological nature of the chiral-symmetric single-particle Hamiltonian can be revealed by the winding number. In the OBC, the real-space winding number of $\hat{H}$ is given by~\cite{PhysRevLett.113.046802}
\begin{equation}
\nu=\frac{1}{L'}\mathrm{Tr}'(\hat{C}\hat{P}[\hat{P},\hat{X}]),
\end{equation}
where the projector $\hat{P}=\sum_{l=1}^L(\ket{\psi_l}\bra{\psi_l}-\hat{C}\ket{\psi_l}\bra{\psi_l}\hat{C}^{-1})$ sums over the lowest half single-particle wave functions $\ket{\psi_l}$ (for the half-filling case), $\hat{C}=\mathbf{I}_L\otimes \hat{\sigma}_x$ is the real-space chiral operator, $\hat{X}$ is the coordinate operator with $X_{j \sigma,j' \sigma'}=\delta_{jj'}\delta_{\sigma \sigma'}$, and $\mathrm{Tr}'$ denotes the trace per volume over the center internal $L'=L/2$ matrix elements which can avoid the boundary effect. For consistency, we also compute the winding number under the PBC $\nu_k$ from the Bloch Hamiltonian (\ref{Hamk}), which can be written as the following off-diagonal form
\begin{equation}\label{hqk}
\hat{H}_\mathrm{B}(k)=\left(\begin{array}{cc}0& q(k)\\ q^\dagger(k)&0\end{array}\right),
\end{equation}
with $q(k)$ an ${a}_{12} \times {a}_{12}$ matrix. The 1D winding number is then obtained by integral over the MBZ~\cite{PhysRevLett.102.196804}
\begin{equation}\label{nu_k}
\nu_k=\frac{1}{2\pi i}\int_0^{\frac{2\pi}{{a}_{12}}} \mathrm{d}k~ \mathrm{Tr}\left[q^{-1}\partial_kq\right].
\end{equation}
The chiral symmetry ensures $\nu_k$ being an integer, which counts the number of times the momentum space Hamiltonian encircles the original point~\cite{PhysRevA.97.052115}. In general, it is not possible to obtain an analytical expression for $\nu_k$ in our model, so numerical integration is employed. Figure~\ref{fig1} (a) shows the topological phase diagram characterized by the real-space winding number $\nu$. The topological phase boundary is consistent with that obtained by the momentum-space winding number $\nu_k$, plotted as the red dashed line.

Due to the gap-closing nature at topological transition points, the localization length $\Lambda$ of the zero-energy mode under the OBC diverges with $\Lambda^{-1}\to0 $ \cite{PhysRevLett.113.046802}. To calculate the localization length of the zero-energy mode,
we can rewrite the Hamiltonian as
\begin{equation}\label{H0}
	\hat{H}_0=\left(\begin{array}{ccccccccc}
		M_1 & T & 0 & 0  & \ldots & 0 & 0 & 0 \\
		T^\dagger & M_2 & T & 0  & \ldots & 0 & 0 & 0 \\
		0 & T^\dagger & M_3 & T  & \ldots & 0 & 0 & 0 \\
		\vdots & \vdots  & \vdots & \vdots & \ddots & \vdots & \vdots & \vdots \\
		0 & 0 & 0 & 0 & \ldots & T^\dagger & M_{L-1} & T \\
		0 & 0 & 0 & 0 & \ldots & 0 & T^\dagger & M_L
	\end{array}\right),
\end{equation}
where $M_j=(m_z+m_{o,j}) \hat{\sigma}_z$ and $T=it_s \hat{\sigma}_y - t\hat{\sigma}_z$.
The zero-energy eigenstate $\psi=\{\psi_1,\psi_2 \cdots \psi_{L-1}, \psi_L \}^T$ with $\psi_j=\{ \psi_{j,\uparrow}, \psi_{j,\downarrow}\}$
can be obtained by solving ${\hat{H}_0}\ket{\psi} =0$ as
\begin{equation}\label{recursion}
	\psi_j=\begin{cases} -T^{-1} M_1 \psi_1 &j=2;\\
		-T^{-1}(T^\dagger \psi_{j-2}+M_{j-1} \psi_{j-1}), &j>2.  \end{cases}
\end{equation}
By setting $\psi_1=\{1,-1\}^T$ as the eigenstate of $\hat{\sigma}_x$, one can obtain $\psi_L$ through Eq. (\ref{recursion}).
Hence, we obtain
\begin{equation}\label{Lambda1}
	\Lambda^{-1}
	=\left|\frac{1}{L}\min \left\{\ln\left|\frac{\psi_{L,\uparrow}}{\psi_{1,\uparrow}}\right|,
	\ln\left|\frac{\psi_{L,\downarrow}}{\psi_{1,\downarrow}}\right|\right\}\right|.
\end{equation}
The numerical result of $\Lambda^{-1}$ is shown in Fig.~\ref{fig1} (b). The divergence of the localization length indicates the topological phase boundary, which is consistent with that revealed by the winding numbers shown in Fig.~\ref{fig1} (a).

Similar to the disorder-induced topological Anderson insulators ~\cite{PhysRevLett.102.136806,PhysRevLett.103.196805}, we find that the commensurate moir\'e potential can drive a topological phase from a trivial insulator, and the finger-shaped elongations of the topological region in Fig.~\ref{fig1} (a) host reentrant topological transitions induced by the moir\'e potential. In Fig.~\ref{fig1} (c), we highlight the reentrant topological transitions by plotting $\nu$ and the single-particle energy gap $\Delta E=E_{L+1}-E_{L}$ under the OBC as functions of $m_o$ with fixed $m_z=2.4$. One can see the sequent trivial-topological-trivial-topological-trivial transition driven by the moir\'e potential strength, with gap closes occurring at each transition point. Such a reentrant topological transition is absent without the moir\'e potential. However, it can be induced by the uniform potential $m_z$ with finite and proper values of $m_o$, such as the topological-trivial-topological-trivial transition shown in Fig.~\ref{fig1} (d). The finite and vanishing values of $\Delta E$ in Figs.~\ref{fig1} (c,d) correspond to the absence and presence of two zero-energy edge modes in the trivial and topological phases under the OBC, respectively. To be more clearly, we show the energy spectrum with respect to the moir\'e potential strength under the OBC in Fig.~\ref{fig1} (e). As the energy gap closes in topological regions, there emerges two exponentially localized states near two ends of the 1D lattice with zero energy, which are shown in Fig.~\ref{fig1} (f). These localized edge states will disappear when the energy gap opens. Note that the reentrant topological phase can be induced by other moir\'e potentials with different choices of ${a}_{1}$ and ${a}_{2}$, and more reentrant transitions can be realized with proper superlattice and Hamiltonian parameters.

\begin{figure}[t!]
\centering
\includegraphics[width=0.48\textwidth]{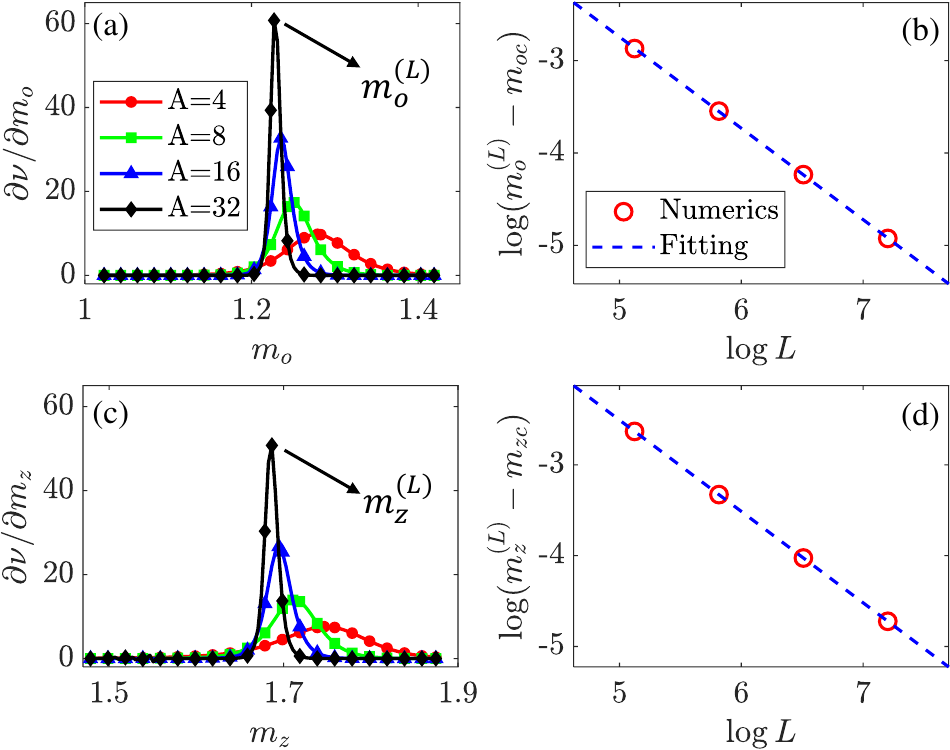}
\caption{Finite-size scaling of the topological invariant. (a) $\partial \nu / \partial m_o$ as a function of $m_o$ with $m_z=2.4$ and various system sizes. The finite-size critical point at $m_o=m_o^{(L)}$ is given by the peak of each curve indicated by the arrow with $m_o^{(L)}$. (b) Finite-site scaling of the distance from $m_o^{(L)}$ to the ideal transition point $m_{oc}=1.223$. (c) $\partial \nu / \partial m_z$ as a function of $m_z$ with $m_o=2.8$ and various system sizes. The peak is indicated by the arrow with $m_z^{(L)}$. (d) Finite-site scaling of the distance from $m_z^{(L)}$ to $m_{zc}=1.678$. Other parameter is $t_s=0.95$.}
\label{fig2}
\end{figure}

We further perform the scale invariance analysis of the real-space winding number in finite-size systems. Note that the finite-size scaling of the fidelity susceptibility and quantum entanglement has been used to explore the critical behaviors in quantum phase transitions ~\cite{PhysRevB.77.245109,PhysRevB.81.064418,Osterloh2002,Zhang_2017}. Figures~\ref{fig2} (a) and \ref{fig2} (c) show the numerical results of $\partial\nu/\partial m_o$ and $\partial \nu/\partial m_z$ as functions of $m_o$ and $m_z$ for various system sizes with fixed $m_z=2.4$ and $m_o=2.8$, respectively. With the increase of $A$ ($L=21A$), the pecks of the curves of $\partial\nu/\partial m_{o,z}$ at $m_{o,z}^{(L)}$ approach to the ideal topological transition points at $m_o=m_{oc}=1.223$ and $m_z=m_{zc}=1.678$ obtained from $\nu_k$. The distances from $m_{o,z}^{(L)}$ to $m_{oc,zc}$ as a function of the system size $L$ are shown in Figs.~\ref{fig2} (b) and \ref{fig2} (d), respectively. We find that they are well fitted by the same power-law scaling form \cite{Chen2016,PhysRevB.95.075116,PhysRevB.110.045119}
\begin{equation}\label{fitting}
	|m_o^{(L)}-m_{oc}|\propto L^\mu,~|m_z^{(L)}-m_{oz}|\propto L^\mu.
\end{equation}
By fitting the curves using Eq.~(\ref{fitting}) in the logarithm form, we obtain the critical exponents $\mu=0.9897$ in Fig.~\ref{fig2} (b) and $\mu=1.0057$ in Fig.~\ref{fig2} (d), respectively. This indicates that these reentrant topological transitions belong to the same universal class with $\mu\approx1$, but different from the that of the disorder-driven topological transitions with
$\mu\approx2$~\cite{PhysRevB.109.L201102}.

\subsection{Robustness against disorders}
\begin{figure}[t!]
\centering
\includegraphics[width=0.48\textwidth]{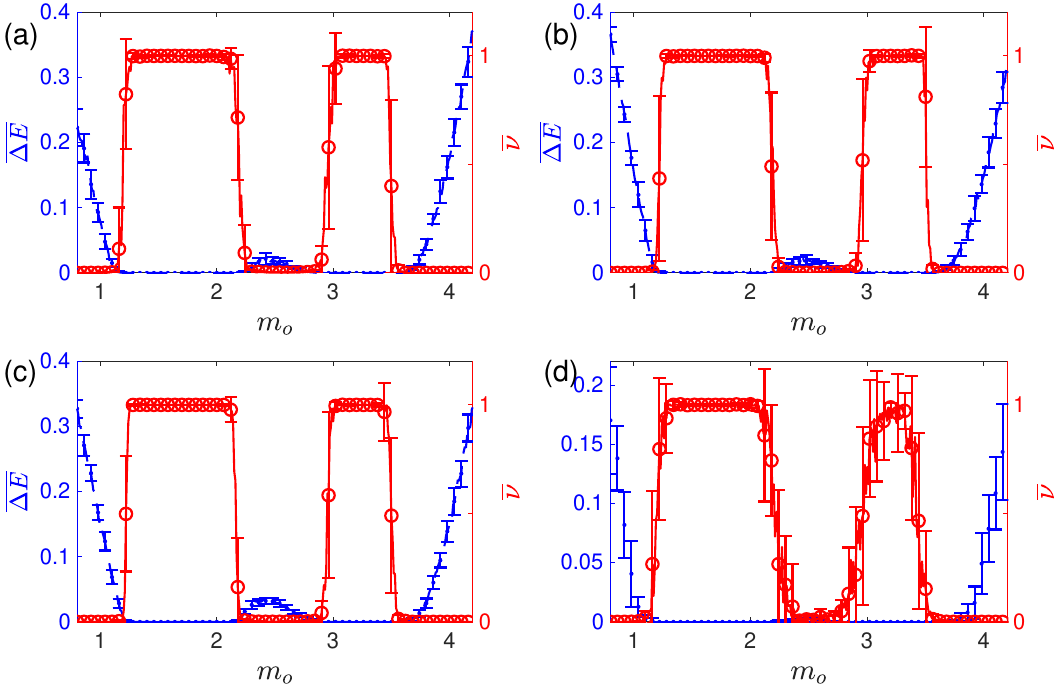}
\caption{Influence of disorders on the reentrant topological transition. Disorder averaged real-space winding number $\overline{\nu}$ and energy gap $\overline{\Delta E}$ are plotted as functions of $m_o$ under the OBC. Disorder is added on the spin-dependent hopping $t_j=t+W_j$ (a), spin-flip hopping $t_{sj}=t_s+W_j$ (b), Zeeman potential $m_{zj}=m_z+W_j$ (c), and all these three components (d). Other parameters are the same as Fig.~\ref{fig1} (c), $W_j\in[-W,W]$ with $W=0.2$, and $20$ different disorder realizations are used. Error bars indicate the standard deviation of the sampled data between different disorder realizations.}
\label{fig3}
\end{figure}

\begin{figure}[t!]
\centering
\includegraphics[width=0.48\textwidth]{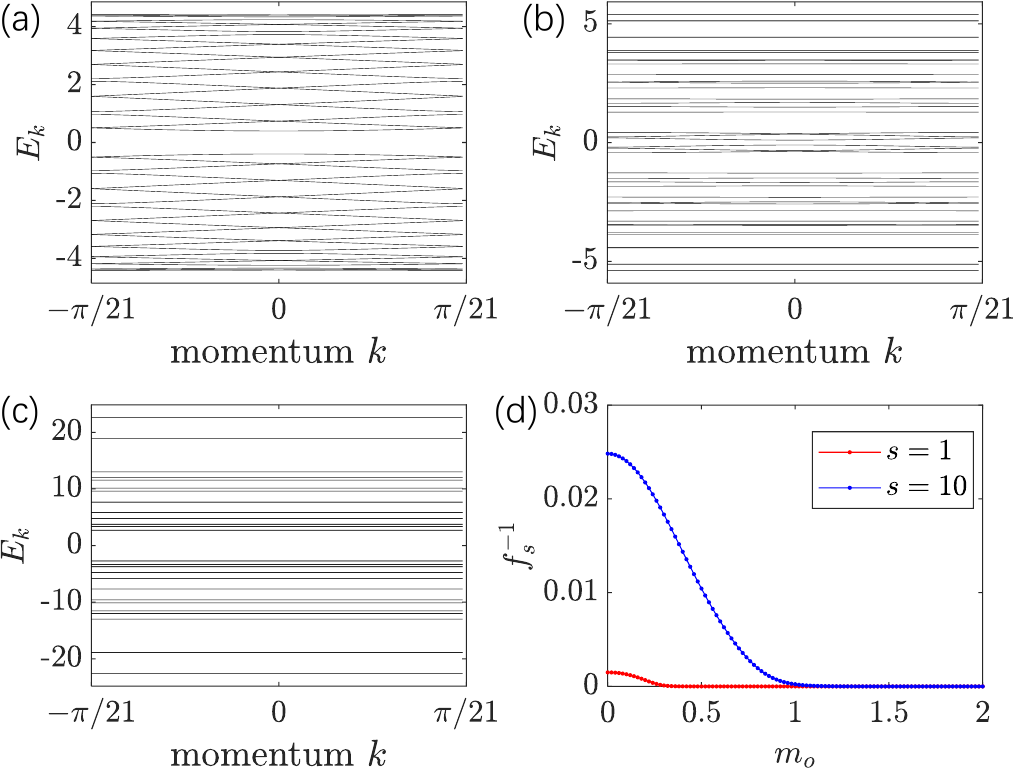}
\caption{Flat band structure and the flatness parameter. The energy spectrum in momentum space for (a) $m_o=0.1$, (b) $m_o=1$ and (c) $m_o=10$ under PBC. (d) Inverse flatness of ${s}$-th energy band $f_{s}^{-1}$ for ${s}=1,10$ as a function of $m_o$. Other parameters are $t_s=0.95$ and $m_z=2.4$.}
\label{fig4}
\end{figure}

As flat-band systems are extremely susceptible to spatial disorder, we provide numerical evidence to ensure that reentrant topological is still robust under disorders. Here, we consider four different situations where disorder is added to the spin-dependent hopping $t_j=t+W_j$, spin-flip hopping $t_{sj}=t_s+W_j$, Zeeman field $m_{zj}=m_z+W_j$, and all of the three parameters. The site dependent disorder $W_j$ is uniformly distributed in $[-W,W]$, where $W=0.2$ is the disorder strength. The averaged real-space winding number
\begin{equation}
\overline{\nu}=\frac{1}{N_s} \sum_i  \nu_i,
\end{equation}
and the averaged energy gap
\begin{equation}
\overline{\Delta E}=\frac{1}{N_s} \sum_i  \Delta E_i,
\end{equation}
where $N_s=20$ disorder realizations are used in our numerical simulation, are calculated to characterize the reentrant topology. In Fig.~\ref{fig3}, we present $\overline{\nu}$ and $\overline{\Delta E}$ as functions of $m_o$ with the same parameters as Fig.~\ref{fig1} (c), where the reentrant phenomenon is clearly observed even for the most stringent situation that all three components are disordered. While further increasing disorder strength $W$, the reentrant region will gradually disappear, indicating the susceptible flat-band structure is more fragile than conventional topological insulators.

\subsection{Localization properties}
\begin{figure}[t!]
\centering
\includegraphics[width=0.48\textwidth]{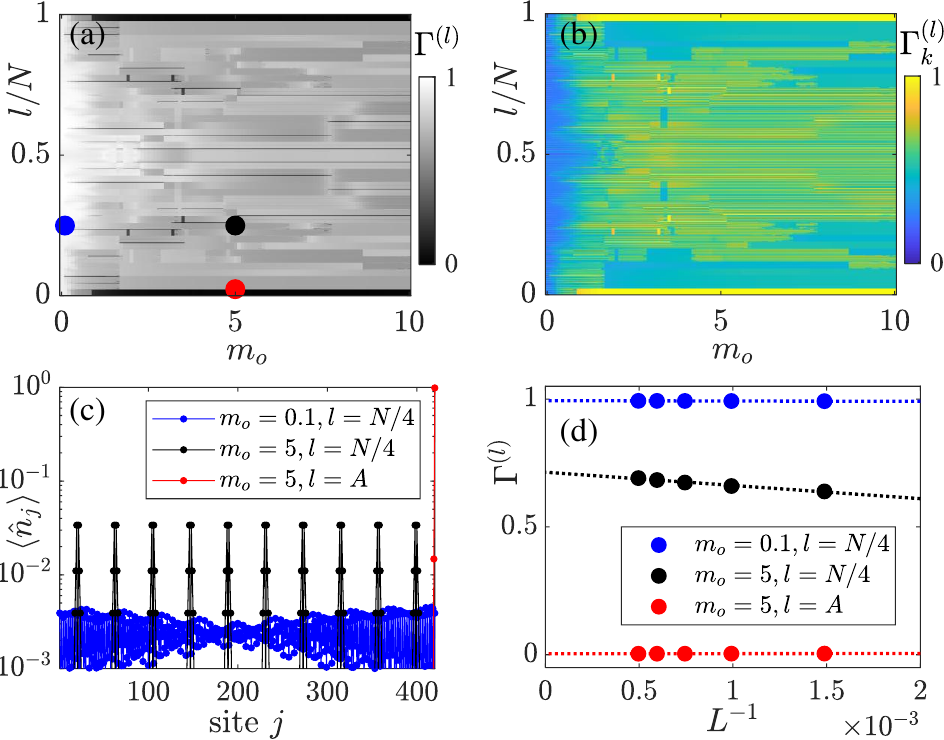}
\caption{Localization properties of single particle eigenstates. (a) Real-space fractal dimension (FD) and (b) momentum-space FD of eigenstates versus $m_o$ with $A=32$. The three colored dots indicate the chosen parameter of the three scaling lines in (d). (c) Density distributions for typical extended (blue), critical (black), and localized (red) states for $A=20$. (d) {\color{black}Finite-size scaling of the real-space FD for three typical states labeled in the legend with $L=21A$, $N=2L$, and $A=\{32,48,64,80,96\}$.} Other parameters are $t_s=0.95$ and $m_z=2.4$.}
\label{fig5}
\end{figure}

We proceed to study the effect of the commensurate moir\'e potential on the localization property of the single-particle eigenstates. Figures~\ref{fig4} (a-c) show the energy bands of $\hat{H}_B(k)$ in the momentum space for $m_o=0.1,1,10$ with fixed $m_z=2.4$, respectively. The results indicate a flattening of the energy bands as $m_o$ increases. We can introduce a dimensionless parameter to characterize the flatness of ${s}$-th band, which is defined as
\begin{equation}
	f_{s}=\frac{\Delta_{s}}{W_{s}}.
\end{equation}
Here $\Delta_{s}=\max \{E_{s}(k)-E_{{s}-1}(k),E_{{s}+1}(k)-E_{s}(k)\}$ denotes the sub-band gap and $W_{s}=\max ||E_{s}(k)-E_{s}(k')||_{k,k'}$ is the bandwidth. When flat band emerges, the bandwidth $W_s$ vanishes while sub-band gap $\Delta_s$ keeps finite, and $f_s$ will tend to infinity. For a better visibility, we plot the inverse of the flatness $f_{s}^{-1}$ as a function of $m_o$ for ${s}=1,10$ sub-bands in Fig.~\ref{fig4} (d). It is clear that the inverse of the flatness tends to zero ($f_s\rightarrow\infty$), and the energy bands become nearly flat under moderate moir\'e potential, which lead to the emergence of compact localized states, similar to those in other moir\'e systems~\cite{PhysRevB.99.075127,PhysRevLett.130.196201,CHEN20233165}.

We adopt the fractal dimension (FD) to reveal the localization properties of eigenstates. For the $l$-th eigenstate, the real-space FD is defined as~\cite{PhysRevLett.83.4590,PhysRevA.103.033325}
\begin{equation}
\Gamma^{(l)}=-\lim_{L\rightarrow\infty}\frac{\ln(IPR^{(l)})}{\ln L},
\end{equation}
where $IPR^{(l)}=\sum_{j,\sigma}|\psi^{(l)}_{j,\sigma}|^4$ is the real-space inverse participation ratio (IPR), and $\psi^{(l)}_{j,\sigma}$ is the probability amplitude of the $l$-th eigenstate. The momentum-space FD is then given as $\Gamma_k^{(l)}=-\lim_{L\rightarrow\infty}\ln(IPR_k^{(l)})/\ln L$, where the momentum $IPR_k^{(l)}$ is obtained by applying a Fourier transformation to the real-space eigenstates. There are extended, localized and critical states in general localization systems. For an extended eigenstate, the wave function is delocalized in the real space with $\Gamma^{(l)}\sim1$, but localized in the dual-momentum space with $\Gamma^{(l)}_k\sim0$. A localized eigenstate is opposite and characterized by $\Gamma^{(l)}\sim0$ and $\Gamma^{(l)}_k\sim1$. In contrast, a critical state is delocalized in both the real and the momentum spaces, and thus has finite FDs $0<\Gamma^{(l)}<1$ and $0<\Gamma^{(l)}_k<1$.

As small energy differences in flat band systems may be rounded off in numerical calculations, we can resolve degenerate energies and corresponding wavefunctions by using translation symmetry operator $\hat{S}$. Let us consider the flat-band Hamiltonian $\hat{H}$ and one of its degenerate subspace $\mathcal{H}$ spanned by the degenerate eigenstates $\hat{V}_{\text{sub}}=\{\ket{\psi^{(1)}},\ket{\psi^{(2)}},\cdots \}$. The translation symmetry operator $\hat{S}$ can be diagonalized within the degenerate subspace since [$\hat{H},\hat{S}]=0$. The projected $\hat{S}$ in the subspace reads
\begin{equation}
	\hat{S}_{\text{sub}}  =  \hat{V}^\dagger_{\text{sub}}   \hat{S}   \hat{V}_{\text{sub}}.
\end{equation}
 Diagonalizing $\hat{S}_{\text{sub}} $
yields its eigenstates $\{\ket{\phi_n}\}$ and eigenvalues $s_n$ as
\begin{equation}
	\hat{S}_{\text{sub}}  \ket{\phi_n} = s_n \ket{\phi_n}.
\end{equation}
Transforming the eigenstate $\ket{\phi_n}$ back to the eigenbasis of $\hat{H}$ by
\begin{equation}
	\ket{\psi_n^\prime} = \sum_l \ket{\psi^{(l)}}\braket{\psi^{(l)} | \phi_n} = \sum_lw_{nl}\ket{\psi^{(l)}},
\end{equation}
with $w_{nl}=\braket{\psi^{(l)} | \phi_n}$, we can verify that $\hat{H}\ket{\psi_n^\prime}=\sum_lE_{d}\ket{\psi^{(l)}}\bra{\psi^{(l)}}\ket{\psi_n^\prime}=E_d\sum_lw_{nl}\ket{\psi^{(l)}}$. Here $E_d$ is the degenerate energy and the updated eigenstates $\{\ket{\psi_n^\prime}\}$ are now both eigenstates of $\hat{H}$ and $\hat{S}$, which ensures the periodicity of $\{\ket{\psi_n^\prime}\}$ under the PBC. Thus, we can use the projected eigenstates $\ket{\psi_n^\prime}$ to calculate localization properties to avoid numerical instability.

Results of $\Gamma^{(l)}$ and $\Gamma_k^{(l)}$ for all OBC eigenstates after translation symmetry projection as functions of $m_o$ for $m_z=2.4$ are shown in Figs.~\ref{fig5} (a) and ~\ref{fig5}(b), respectively. As $m_o$ increases from 0 to a moderate value $m_o\sim1$, a considerable part of eigenstates crossover to the delocalized critical states from the extended states. For moderate $m_o$, the extended, critical, and localized states coexists in the system. For sufficiently large $m_o$, only critical and localized states are exhibited. The density distributions $\braket{\hat{n}_{j}}=\braket{\hat{n}_{j,\uparrow}}+\braket{\hat{n}_{j,\downarrow}}$ of three distinct eigenstates are shown in Fig.~\ref{fig5} (c). One can observe that the delocalized critical states in the commensurate lattice exhibit translation invariance with respect to the supercells, which reflect the moir\'e patten of the superlattice. In Fig.~\ref{fig5} (d), we perform the finite-size scaling of the real-space FD and confirm the three distinct localization states. Note that the scaling analysis is taken by choosing the eigenstate with the fixed index $l/N$ for different system sizes, such as $l/N=1/42$ and $l/N=1/4$ for the localized and critical states in Fig.~\ref{fig5} (d), respectively. The reason for this choice is that eigenstates with the same index $l/N$ share similar FD structures and density distributions.

\subsection{Periodic-moir\'{e}-spin density waves}
\begin{figure}[t!]
\centering
\includegraphics[width=0.48\textwidth]{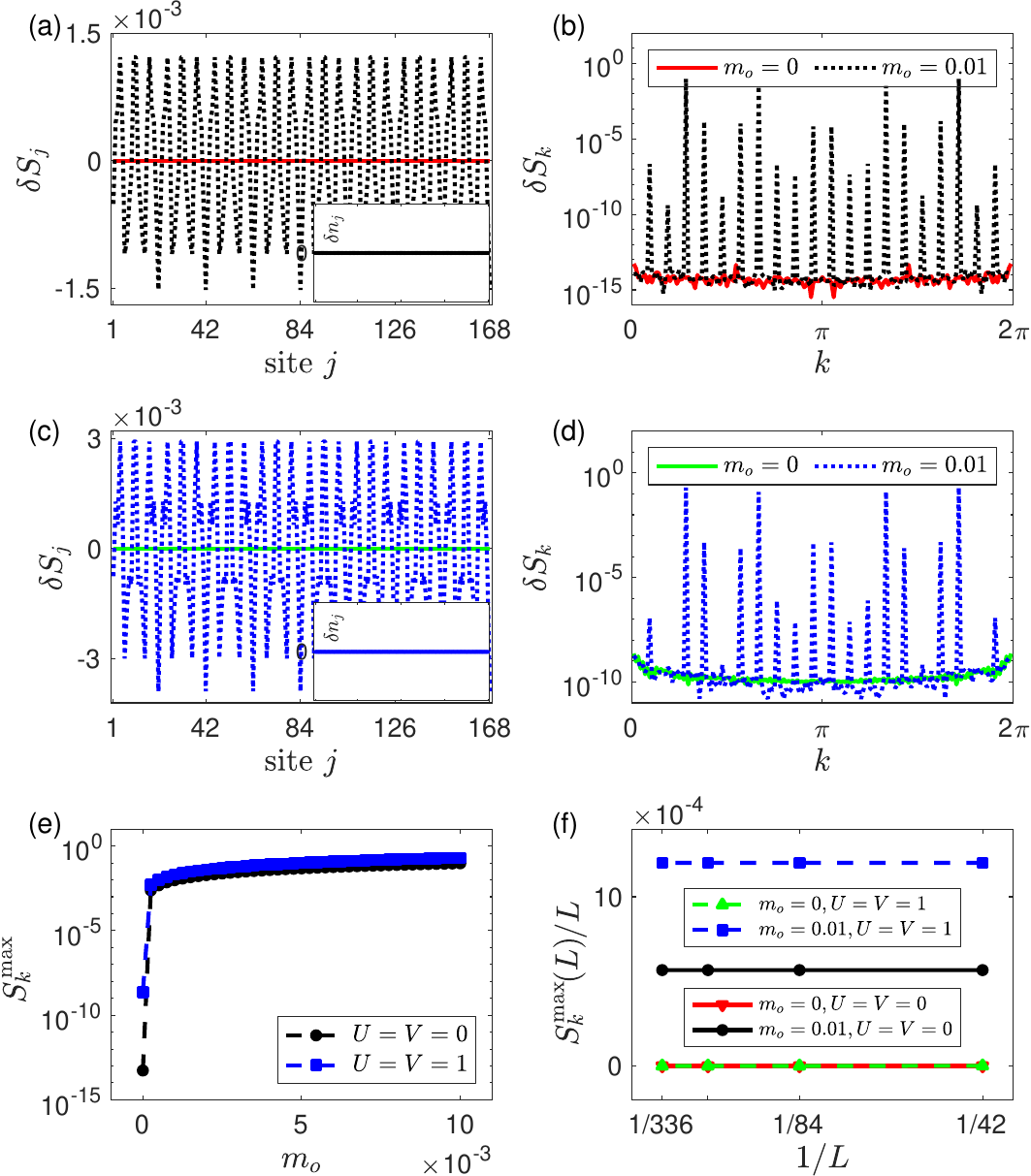}
\caption{Ground state spin density wave order. (a,c) The real-space spin density fluctuation ${\delta S_j}$ plotted for non-interacting case with $U=V=0$ (a), and interacting case with $U=V=1$ (c). Solid lines for $m_o=0$  and dashed curves for $m_o=0.01$. Inset plots are the charge density fluctuation ${\delta n_j}$ for $m_o=0.01$ where no CDW occurs. (b,d) The momentum-space spin density fluctuation ${\delta S_k}$ for the non-interacting and interacting cases in (a,c), respectively. (e) The maximum local spin density ${S_k}^\mathrm{max}$ as functions of $m_o$ for non-interacting (black circles) and interacting (blue squares) cases. (f) The scaling of maximum local spin density versus system size $L=21A$ with $A=2,4,8,16$ for the non-interacting (solid lines) and interacting (dashed lines) cases. Other parameters are $t_s=0.95$ and $m_z=2.4$.}
\label{fig6}
\end{figure}

We first report the emergent of periodic-moir\'e-spin density wave (PM-SDW) order in the Mott insulating phase driven by the moir\'e potential $m_o$, and then discuss the influence of the on-site interaction $U$ and nearest-neighbor interaction $V$. To characterize the PM-SDW order, we calculate the spin density fluctuation
\begin{equation}\label{eqsdf}
{\delta S_j}={S_j}-\bar{{S}}
\end{equation}
in the real space. Here ${S_j}=\braket{\hat{n}_{j,\uparrow}}-\braket{\hat{n}_{j,\downarrow}}$ is spin density on site $j$ and $\bar{{S}}$ is the spatially averaged spin density. The local spin density in the momentum space is given by ${S_k}=\sum_j{S_j}\cos(jk+\phi_0)$ \cite{PhysRevB.103.085136}, where $k=2\pi {i}A/L$ (${i}=1,2,\cdots,{a}_{12}$) is the wave vector and $\phi_0$ denotes an arbitrary phase. The spin densities may subsequently become modulated by the moir\'e superlattice, revealing a PM-SDW period that is commensurate with the supercell of $a_{12}=21$ sites. Thus, one can extract these properties from the peaks of the density distribution by transforming ${\delta S_j}$ to the momentum space ${\delta S_k}$. The density distribution of the PM-SDW state shows a period of ${a}_{12}$ sites in the real space, which corresponds to ${a}_{12}-1$ typical peaks in the momentum space, as the peak at $k=0$ is removed by subtracting the average spin density $\bar{{S}}$ in Eq.~(\ref{eqsdf}). 

In Figs.~\ref{fig6} (a,b), we show the spin density fluctuation for $m_o=0,0.01$ in the real and momentum spaces, respectively. The periodicity of the spin-density wave (SDW) emerges when $m_o$ changes from $0$ to $0.01$. These are ${a}_{12}-1=20$ peaks in the spin density fluctuation ${S_k}$, which clearly reveal the PM-SDW nature of the many-body ground state induced by the moir\'e potential. The charge density fluctuation ${\delta n_j}=\braket{\hat{n}_{j}}-\bar{{n}}$ is also plotted as an inset figure, which indicates the absence of the charge-density wave (CDW) in this non-interacting case. Similar results for the interacting case with $U=V=1$ are shown in Figs.\ref{fig6} (c,d). Although the DMRG sweeps convergent to our error goal, two peaks near the magnitude of $10^{-10}$ are indistinguishable in the background due to the approximate same magnitude numerical errors in DMRG cutoffs. The maximum local spin density $S_{k}^{\mathrm{max}}$ extracted from the highest peak, is plotted as functions of $m_o$ in Fig.~\ref{fig6} (e). These numerical results indicate that the emergence of the PM-SDW order for both non-interacting and interacting cases is due to the moir\'e superlattice potential. Moreover, a very small moir\'e potential strength is sufficient to drive the PM-SDW order. We also consider the dependence of the PM-SDW on the system size and reveal
the scaling of the maximum local spin density $S_{k}^{\mathrm{max}}$ in Fig.~\ref{fig6} (f). For the ordinary Mott insulator without SDW order, when $m_o=0$ the local spin densities keep vanishing for all system sizes. For finite $m_o$, the maximum local spin density per site is non-vanishing and independent on the system size, which indicate that the SDW order is preserved in the thermodynamic limit. In cold-atom systems, the SDW order could be experimentally detected using the spin-polarized scanning tunneling microscopy~\cite{RevModPhys.81.1495}.

\begin{figure}[t!]
\centering
\includegraphics[width=0.48\textwidth]{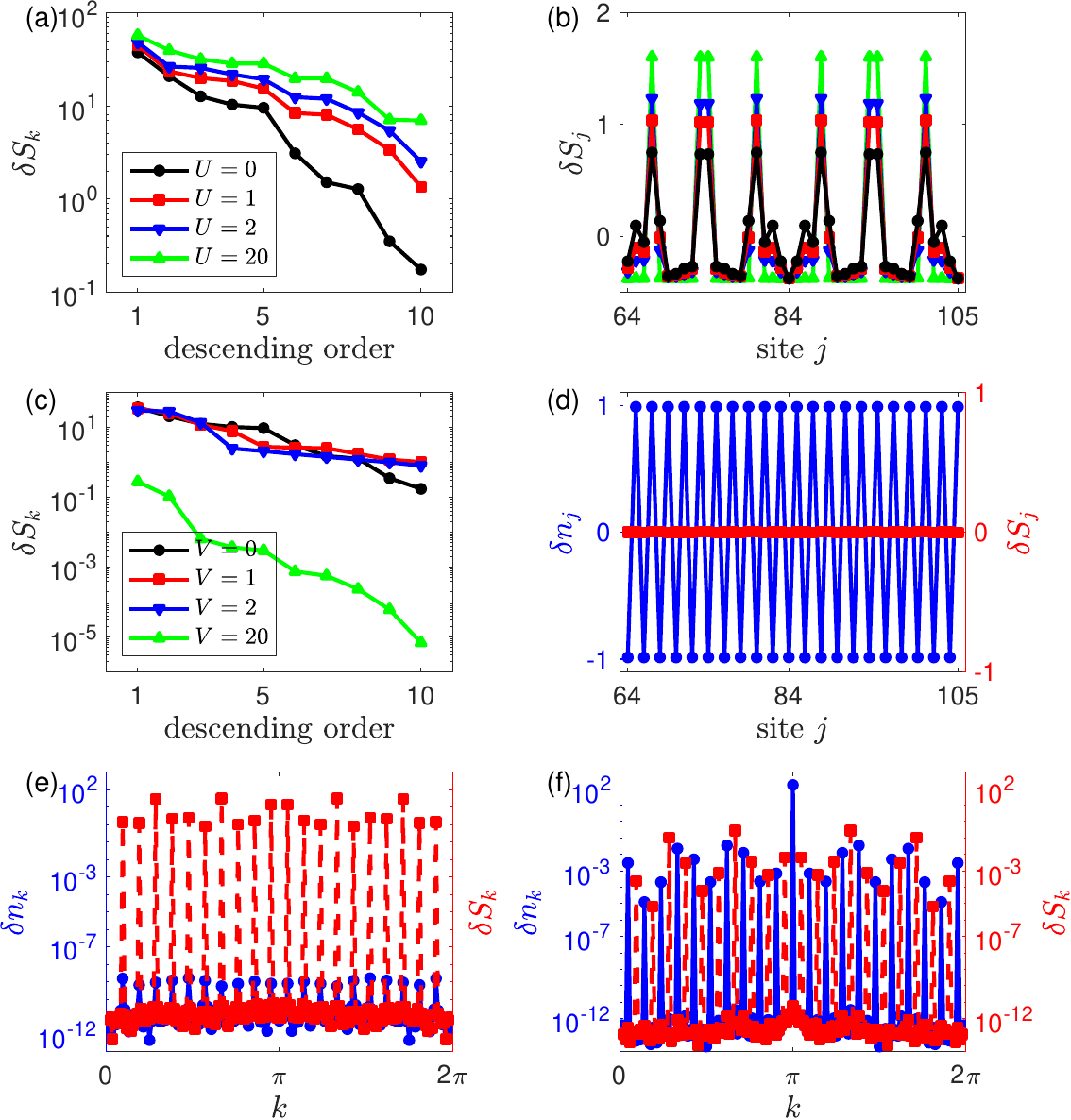}
\caption{Influence of two interactions on spin and charge density wave orders. (a) Peak values of the momentum-space spin density fluctuation ${\delta S_k}$ in region $[0, \pi)$ plotted in descending order for several $U$s with $V=0$. (b) Distributions of the real-space spin density fluctuation ${\delta S_j}$ in the two centered supercells. (a) and (b) share the same legend shown in (a). (c) Peak values of ${\delta S_k}$ plotted in descending order for several $V$s with $U=0$. (d) The real-space charge density fluctuation ${\delta n_j}$ and ${\delta S_j}$ for $V=20$. (e,f) Distributions of the momentum-space charge density fluctuation ${\delta n_k}$ and ${\delta S_k}$ for $V=2$ (e) and $V=20$ (f). Other parameters are $t_s=0.95$, $m_z=2.4$ and $m_o=2$.}
\label{fig7}
\end{figure}

We further investigate the effects of the on-site and nearest-neighbor interactions on the PM-SDW. For this purpose, we fix the moir\'e potential strength at $m_o=2$, where the many-body ground state exhibits the PM-SDW order even in the non-interacting limit. In the presence of an on-site interaction $U$, particles with spin imbalances have lower energy and thus the SDW order is enhanced by increasing $U$. We observe the positive effect of on-site interaction on the PM-SDW from  the local spin densities in the momentum space ${\delta S_k}$ and its real-space counterpart ${\delta S_j}$, as shown in Figs.~\ref{fig7} (a) and ~\ref{fig7} (b), respectively. Note that all peak values of ${\delta S_k}$ are two-fold degenerate due to the reflection symmetry with respect to $k=\pi$. In Fig.~\ref{fig7} (a), we consider the ten peaks in region $k\in[0, \pi)$ and depict their values in descending order for several $U$s. Peak values of ${\delta S_k}$ keep rising as $U$ increases, which demonstrates the enhancement of the PM-SDW order. The corresponding real-space fluctuation ${\delta S_j}$ in the two centered supercells is plotted in Fig.~\ref{fig7} (b), which shows the same moir\'e periodicity and increased amplitudes of the SDWs as $U$ increases.

Strong nearest-neighbor interaction $V$ may suppress the SDW and induce the CDW, since the spin imbalance has lower interaction energy in this case. In Fig.~\ref{fig7} (c), 20 peak values of ${\delta S_k}$ are plotted in the descending order for $V=0,1,2,20$. The PM-SDWs are almost unchanged for small to moderate $V$s, but are inhibited for large $V$. We focus on the strong nearest-neighbor interaction case with $V=20$, and depict ${\delta n_j}$ and ${\delta S_j}$ in the middle two supercells in Fig.~\ref{fig7} (d). The CDW with a period of two sites emerges in this case, while the SDW almost vanishes. For a better comparison, we present the momentum-space counterparts ${\delta n_k}$ and ${\delta S_k}$ for $V=2$ and $V=20$ in Figs.~\ref{fig7} (e,f), respectively. For $V=2$ in Fig.~\ref{fig7} (e), the PM-SDW  dominates and a minor PM-CDW coexists with very small magnitudes.
For $V=20$ in Fig.~\ref{fig7} (f), the PM-SDW is strongly suppressed with average spin density $\bar{S}=0$, and the CDW dominates with the average charge density $\bar{{n}}=1$. Here the CDW order shows the dominated wave number at $k=\pi$. The moir\'e periodicity can still be revealed from the 20 peaks for PM-CDW or PM-SDW, which are three orders of magnitude smaller than that for the CDW order at $k=\pi$.

\subsection{Interacting topological phases}
\begin{figure}[t!]
\centering
\includegraphics[width=0.48\textwidth]{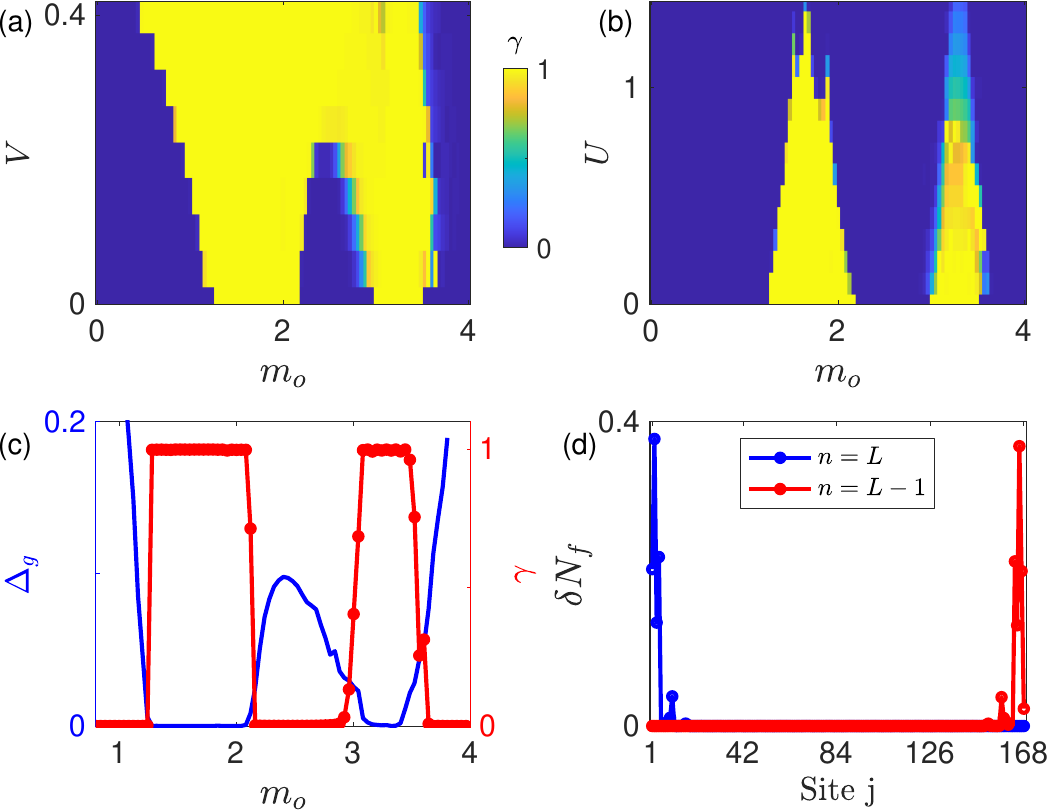}
\caption{Topological phase diagram and related properties in the interacting region. (a,b) Many-body Berry phase $\gamma$ in units of $\pi$ plotted in the $m_o$-$V$ plane with $U=0$ (a) and in the $m_o$-$U$ plane with $V=0$ (b). (c) Excitation gap $\Delta_g$ under the OBC and $\gamma$ as functions of $m_o$ for $U=0.1$. (d) Distribution of two zero-energy excitation modes near half filling $\delta N_f$ for $U=0.1$ and $m_o=1.7$. Other parameters are $t_s=0.95$, $m_z=2.4$, $A=2$ for $\gamma$, and $A=8$ for others.}
\label{fig8}
\end{figure}

In the interacting case, the topology of the ground state at half filling can be characterized by the many-body Berry phase under the twisted, which is given by PBC~\cite{RevModPhys.82.1959,PhysRevB.84.195107,PhysRevX.7.031057}
\begin{equation}
\gamma=\frac{1}{\pi}\oint \mathrm{d}\theta\braket{\Psi^g(\theta)|i\partial_\theta|\Psi^g(\theta)} \ \mathrm{mod}\ 2
\end{equation}
in units of $\pi$. Here $\ket{\Psi^g(\theta)}$ is the many-body ground state of $L$ fermions with the twist phase $\theta$. The quantized Berry phase is $\gamma=1$ for the topological phase, while $\gamma=0$ for the trivial phase. In Figs.~\ref{fig8} (a,b), we show the numerical results of $\gamma$ in the $m_o$-$V$ and $m_o$-$U$ planes for $m_z=2.4$, respectively. The reentrant topological phase is preserved for certain $U$s or $V$s, before the on-site or nearest-neighbor interactions are dominated. It's worth emphasizing that the reentrant topological phase is in the parameter region of the PM-SDW, as discussed previously. Thus, the topological PM-SDW can be exhibited for interacting fermions in the moir\'e superlattice.

In the interacting topological phase under the OBC, the ground states near half filling are two-fold degenerate with zero-energy excitations localized near two edge of the lattice. We numerically compute the excitation gap $\Delta_g$ under the OBC, which is defined as the energy gap between the first excited state and the ground state. The excitation gap and the corresponding Berry phase as functions of $m_o$ for $U=0.1$ are plotted in Fig.~\ref{fig8} (c). The gap closing and reopening behavior is consistent with that of the many-body Berry phase. To show the edge excitations in the topological phase, we compute the density distribution
\begin{equation}
\delta N_f (j)=\braket{\Psi^g_{N_f+1}|\hat{n}_j|\Psi^g_{N_f+1}}-\braket{\Psi^g_{N_f}|\hat{n}_j|\Psi^g_{N_f}},
\end{equation}
which is defined as the difference of the ground-state distributions between $N_f+1$-filling and $N_f$-filling with $N_f=L$ or $L-1$. This corresponds to adding or removing one quasiparticle to the half-filling ground state. In the topological phase, the added (removed) quasiparticle tends to localize near one end of the 1D lattice, as shown in Fig.~\ref{fig8} (d).

\section{\label{secf}Discussion}

\begin{figure}[t!]
\centering
\includegraphics[width=0.48\textwidth]{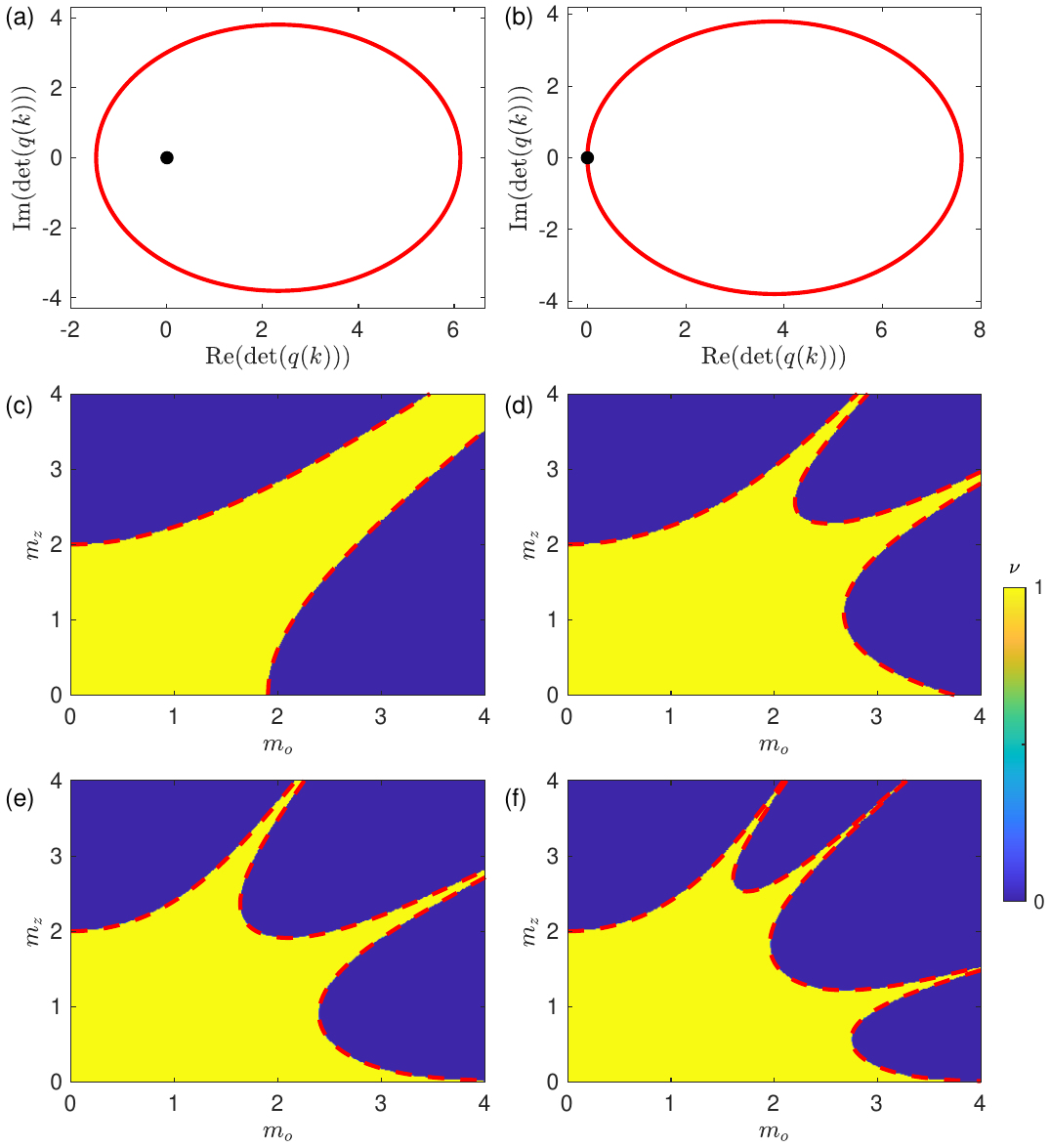}
\caption{The loop of winding number and additional phase diagrams. (a) The loop of $\det(q(k))$ winds around the origin (black dot) in the topological region for $a_1=2$, $m_z=2.4$ and $m_o=1.8$ as the wavenumber $k$ runs through the reduced Brillouin zone. (b) The loop of $\det(q(k))$ passes though the origin (black dot) at at the critical point for $a_1=2$, $m_z=2.4$ and $m_o=\sqrt{1.76}$. Momentum-space winding number $\nu$ in the $m_o$-$m_z$ plane for (c) $a_1=2$, $t=1$, and $t_s=0.95$; (d) $t=t_s=1$ with $(a_1,a_2)=(3,7)$; (e) $t=t_s=1$ with $(a_1,a_2)=(2,5)$; and (f) $t=t_s=1$ with $(a_1,a_2)=(3,5)$. Red dashed curves in (c-f) are phase boundaries solved from the renormalized Zeeman strength.}
\label{fig9}
\end{figure}

Before concluding, we discuss the underlying mechanism of the reentrant topological transition and show that this phenomenon is generic under other moir\'e potentials. The reentrant topological transition can be characterized by the renormalization of the Zeeman field $m_z$ by the moir\'e modulation $m_o$. To reveal this point, we can start with a simple potential  $m_{o,j}=m_o\cos(2\pi j/a_1)$ with $a_1=2$ (and $a_2=\infty$), such that the analytical solution of the topological phase boundaries can be obtained. In this simple case, the momentum-space off-diagonal Hamiltonian (\ref{hqk}) is a $4\times4$ matrix with off-diagonal matrix $q(k)=-2m_z-4[t\cos(k)-it_s\sin(k)]\sigma_z-2m_o\sigma_x$. The winding number equals to how many times $\det(q(k))$ encircles the origin point when the wavenumber $k$ sweeps though the reduced Brillouin zone $[0,2\pi/a_1)$. The topological transition point corresponds to the condition satisfying $\det(q(k))=0$. Fig.~\ref{fig9} (a) shows that $\det(q(k))$ encircles the origin point for the topological phase, while Fig.~\ref{fig9} (b) represents the critical case where the loop of $\det(q(k))$ passes through the origin point. The imaginary part $\mathrm{Im}(\det(q(k)))=32tt_s\cos(k)\sin(k)=0$ reveals that the solution occurs at $k=0$ and $k=\pi/2$. For the real part $\mathrm{Re}(\det(q(k)))=0$, we obtain two equations for $k=0$ and $k=\pi/2$, respectively
\begin{align}
m_z^2-m_o^2-4t^2=0,\\
m_z^2-m_o^2+4t_s^2=0.
\end{align}
Thus, the topological phase boundaries satisfy the following equations
\begin{align}
\tilde{m}_z^2 \equiv m_z^2-m_o^2=4t^2,\\
\tilde{m}_z^2 \equiv m_o^2-m_z^2=4t_s^2,
\end{align}
where $\tilde{m}_z$ denotes the renormalized Zeeman field as a function of the moir\'e modulation $m_o$. The topological transition occurs at $\tilde{m}_z=\pm 2t$ and $\tilde{m}_z=\pm 2t_s$. For $t=1$ and $t_s=0.95$ shown in Fig.~\ref{fig9} (c), the topological transition points in the phase diagram are given by $m_z=\sqrt{m_o^2+4}$ and $m_z=\sqrt{m_o^2-3.61}$, which consistent with the numerical results. 

Although the reentrant topological transition is absent in this simple case, it can be exhibited  under the renormalization from the moir\'e potentials of other proper values of $\{a_1, a_2\}$.
The analytical method presented above is reliable by using the condition equations $\det(q(0))=0$ and $\det(q(\pi/a_{12}))=0$. For $a_{12}=3\times7=21$ considered previously, they correspond to two 21-th order functions with a maximum of $42$ real roots, such that the analytical expressions of the phase boundaries can not be obtained. However, we can still derive the renormalization relation between $m_z$ and $m_o$ when $t_s\approx t$. The localization length and the recursive relation of the zero-energy eigenstate for Eq. (\ref{H0}) reads
\begin{equation}
	T^\dagger \psi_{j-1}^T + M_j \psi_{j}^T + T \psi_{j+1}^T = 0,~~~j>1,
\end{equation}
which can be expressed as
\begin{align}
	\begin{split}
		-t\psi_{j-1,\uparrow} -t_s\psi_{j-1,\downarrow}+(m_z+m_{o,j}) \psi_{j,\uparrow}&\\
		-t\psi_{j+1,\uparrow} +t_s\psi_{j+1,\downarrow}&=0,
	\end{split}\\
	\begin{split}
		t\psi_{j-1,\uparrow} +t_s\psi_{j-1,\downarrow} -(m_z+m_{o,j}) \psi_{j,\downarrow}&\\
		-t\psi_{j+1,\uparrow} +t_s\psi_{j+1,\downarrow}&=0.
	\end{split}
\end{align}
With $\psi_{1}=\{1,-1\}$ and the approximation $t_s\approx t$, the two equations can be solved by the following recursive relation
\begin{equation}
	\psi_{j,\uparrow}=-\psi_{j,\downarrow}=\frac{m_z+m_{o,j-1}}{2t} \psi_{j-1,\uparrow},~~~j>1.
\end{equation}
One can derive that
\begin{equation} \label{psi_L}
	\psi_{L}=\prod_{j=1}^{L-1} \frac{m_z+m_{o,j}}{2t} \psi_{1}.
\end{equation}
By substituting Eq. (\ref{psi_L}) into Eq. (\ref{Lambda1}),  we obtain the the localization length in the thermodynamic limit as
\begin{equation}
	\begin{split}
		\Lambda^{-1}
		&=\left|\lim_{L\to\infty}  \frac{1}{L} \ln  \prod_{j=1}^{L-1} \left|  \frac{m_z+m_{o,j}}{2t}\right| \right|\\
		&=\left|\lim_{L\to\infty}  \frac{1}{L} \left(\ln  \prod_{j=1}^{L} \left|  \frac{m_z+m_{o,j}}{2t}\right| -\ln \left|\frac{m_z+m_{o,L}}{2t}\right|\right) \right|\\
		&=\left|\frac{1}{a_{12}}\ln \prod_{j=1}^{a_{12}} \left|{m_z+m_{o,j}}\right| -\ln|2t| \right|\\
		&=\left|\ln |\tilde{m}_z| -\ln|2t| \right|,
	\end{split}
\end{equation}
where $L=Aa_{12}$ with $A\to\infty$. As $\lim_{L\to\infty}( \ln |m_z+m_{o,L}|-\ln|2t|)/L\to 0$ for finite Hamiltonian parameters, we obtain the renormalized Zeeman stength 
\begin{equation}
	\tilde{m}_z=\prod_{j=1}^{a_{12}} ({m_z+m_{o,j}})^{\frac{1}{a_{12}}},
\end{equation}
as a function the moir\'e modulation. The phase boundary can be obtained by solving  $\tilde{m}_z=\pm 2t$, which are two $a_{12}$-th order equations with a maximum of $2a_{12}$ real roots, which indicates the possible reentrant transitions between trivial and topological phases. In Fig.~\ref{fig9} (d), we show the topological phase boundaries solved from the renormalization relation $\tilde{m}_z=2t$ for the case of $\{a_1=3,a_2=7\}$ (see Figs.~\ref{fig9} (e,f) for other cases), consistent with those in Fig.~\ref{fig1} (a). Moreover, we numerically show that the reentrant topological transition is generic under other moir\'e potentials, such as $\{a_1,a_2\}=\{2,5\},\{3,5\}$ in Figs.~\ref{fig9} (e) and ~\ref{fig9}(f), respectively.

\section{Conclusions}
To summarize, we have demonstrated that the 1D moir\'e potential can induce trivial-topological-trivial-topological-trivial multiple transitions in both single-particle and many-body regions in a spin-$1/2$ fermionic optical superlattice. We have uncovered the topological phases with zero-energy edge modes or excitations and nontrivial topological numbers. The scaling exponents of topological transitions for both moir\'e and uniform Zeeman potentials have been revealed and agree with each others. We have also investigated the localization property with the moir\'e-induced nearly flat bands and delocalized critical states. The topology and localization in our moir\'e superlattices are different from those in disorder-induced TAIs. Then, we have unveiled the PM-SDW orders of the many-body ground state, which are instantly induced by turning on the moir\'e potential. The on-site interaction can enhance the PM-SDW, while a sufficient nearest-neighbor interaction suppresses the SDW and induces the CDW. Finally, we have generalized our findings to the interacting region by means of the DMRG method. The reentrant topological phase persists for a finite interaction strength, after which two topological regions either merge or vanish, as the nearest-neighbor and on-site interactions enhance and destroy the topology, respectively. {\color{black}The proposed model could be realized in future experiments of ultracold atoms with properly engineered Raman optical lattices and interatomic interactions.} It would also be interesting to further explore reentrant topological phases and SDW orders in 2D and incommensurate moir\'e systems.

\section*{Methods}
The topological and localization properties of the system in the single-particle case are directly obtained from the real-space Hamiltonian $\hat{H}$ in Eq. (\ref{Ham}) and the momentum-space Hamiltonian $\hat{H}_\mathrm{B}(k)$ in Eq. (\ref{Hamk}). Both Hamiltonians are constructed using Matlab, and the real-space and momentum-space physical quantities are numerically calculated by the eigen method of these matrices using MATLAB version R2023b. The physical quantities of half-filling many-body ground states in the noninteracting limit are obtained from the eigen method by summing over the lowest lying half single-particle physical quantities.
As a supercell has $21$ sites, the eigen method is far beyond the availability with more than one supercell in the many-body interacting case. The many-body ground states at half-filling in the presence of the on-site and nearest-neighbor interactions are simulated by the DMRG method with matrix-product state representation. We use the itensor library~\cite{itensor} in our numerical simulations. In the simulation of the many-body Berry phase, the system size $L=21A$ with $A=2$ is considered, the bond dimension of the maximum matrix product states is set to $200$, and $24$ DMRG sweeps or $10^{-6}$ relative energy error goal are taken. These criteria are sufficient in our simulations as the topological invariant is robust against disorders and finite-site effect. In critical regions, the DMRG method may struggle to converge, even with a larger bond dimension, more DMRG sweeps, or a smaller relative energy goal. This is due to narrow energy gaps in these regions, which lead to a not strictly quantized Berry phase. In the calculation of other quantities, we increase the supercell number to $A=8,16$ ($L=168,336$ with $N_a=L$ spin-1/2 fermions), with up to $400$ bond states and $36$ DMRG sweeps or $10^{-8}$ relative energy error goal to achieve convergent results.

\section*{Data availability}
Data underlying the results presented in this paper are available in figshare with the identifier \href{https://doi.org/10.6084/m9.figshare.29264174.v1}{doi: 10.6084/m9.figshare.29264174}.

\section*{Code availability}
The computer codes used in this paper are available from the corresponding authors upon reasonable request.

\bibliography{reference}

\begin{acknowledgments}
This work was supported by the National Key Research and Development Program of China (Grants No. 2022YFA1405304 and No. 2024YFA1409300), the National Natural Science Foundation of China (Grants No. 12174126 and No. 12104166), the Guangdong Basic and Applied Basic Research Foundation (Grant No. 2024B1515020018), the Science and Technology Program of Guangzhou (Grant No. 2024A04J3004), and the Open Fund of Key Laboratory of Atomic and Subatomic Structure and Quantum Control (Ministry of Education).
\end{acknowledgments}

\section*{Author contributions}
Guo-Qing Zhang performed the many-body calculations and wrote the first draft, Ling-Zhi Tang conceived and performed the single-particle part, L.F. Quezada analyzed the data,  Dan-Wei Zhang conceived the initial idea,  Shi-Hai Dong and Dan-Wei Zhang jointly supervised the project. All authors discussed the results and revised the manuscript.

\section*{Competing interests}
The authors declare no competing interests.

\end{document}